\documentclass[useAMS,usenatbib]{mn2e}
\usepackage{graphicx}
\usepackage{amssymb}
\title[Searching for strong galaxy-galaxy lenses]{A method to search for strong galaxy-galaxy lenses in optical imaging surveys}
\author[Kubo \& Dell'Antonio]{Jeffrey M. Kubo$^{1,2}$\thanks{E-mail:
kubo@fnal.gov} and Ian P. Dell'Antonio$^{2}$\\
$^{1}$Centre for Particle Astrophysics, Fermi National Accelerator Laboratory, MS 127, P.O. Box 500, Batavia, IL 60510\\
$^{2}$Physics Department, Brown University, Box 1843, Providence, RI 02912}
\begin{document}



\maketitle

\label{firstpage}

\begin{abstract}
We present a semi-automated method to search for strong galaxy-galaxy lenses in optical imaging surveys.  Our search technique constrains the shape of strongly lensed galaxies (or arcs) in a multi-parameter space, which includes the third order (octopole) moments of objects.  This method is applied to the Deep Lens Survey (DLS), a deep ground based weak lensing survey imaging to $R\sim26$. The parameter space of arcs in the DLS is simulated using real galaxies extracted from deep HST fields in order to more accurately reproduce the properties of arcs.  Arcs are detected in the DLS using a pixel thresholding method and candidate arcs are selected within this multi-parameter space.  Examples of strong galaxy-galaxy lens candidates discovered in the DLS F2 field (4 square degrees) are presented.
\end{abstract}


\section{Introduction}
\label{sec:introduction}

Strong gravitational lensing is a powerful method with which to study the mass and mass distribution of foreground lens galaxies, independent of dynamical assumptions.  The majority of confirmed galaxy scale lenses are still galaxy-quasar lenses \citep{falco05}, though the number of galaxy-galaxy lenses is quickly catching up (i.e. \citealt{bolton07}).  Recent work in gravitational lens modeling suggests that galaxy-galaxy lens systems, where the source is extended (a galaxy), provide the most accurate lens models \citep{warren03}.  Galaxy-galaxy lens systems can also be used to study the lensed galaxies themselves, providing a window into higher redshift galaxies \citep{allam07}.  With future wide field surveys such as the Dark Energy Survey (DES) \citep{annis05} or LSST \citep{tyson02}, large samples of strong galaxy-galaxy lenses could also be used to put constraints on cosmology through the statistics of strong lensing \citep{kochanek96} \citep{chae02} \citep{linder04}.  

Miralda-Escude \& Lehar (1992) were the first to point out that a large number of Einstein rings (galaxy scale lens systems in which the source is a galaxy) should exist in the optical.  They pointed out that the detection of rings depends on (1) angular resolution (2) the ability to identify rings among the large number of optical sources.  They suggested looking for likely lenses (massive ellipticals) and subtracting out the light from the foreground lens galaxy, as a method of detecting candidate ring systems.  Many systems have since been detected in the optical, but mostly through visual selection.  Examples include the four systems in the HST Medium deep survey \citep{ratnatunga99}, one system in the Ultra Deep Field \citep{blakeslee04}, and the three systems in the AEGIS survey \citep{moustakas07}.  The GOODS survey \citep{fassnacht04} has discovered a number of candidate systems using a method similar to that suggested by Miralda-Escude \& Lehar (1992).  Here image subtraction in GOODS works well because of the HST resolution in all filters and relatively small survey area.  Much progress has been made recently in finding strong galaxy-galaxy lensed systems in the SDSS redshift survey \citep{bolton04}.  Here the emission line(s) of lensed source galaxies are detected in the spectrum of foreground lens galaxies.  This search method is quite powerful but requires galaxy spectra, and therefore is not applicable to purely imaging surveys.  
  
Large area ground based optical surveys, especially weak lensing surveys, should be excellent places to search for candidate strong galaxy-galaxy lens systems.  These surveys are wide area, deep, and because of the observational requirements to measure the weak lensing signal, have relatively good resolution (seeing).  Good seeing is essential in separating the arc from the host lens galaxy, and the wide area increases the number of possible systems.  Since weak lensing surveys are also focused on the detection of galaxy clusters, from maps of weak lensing shear \citep{kaiser93} or optical cluster finding \citep{hansen05}, structure within the survey fields is well understood.  Thus the environments of strong lens systems, which can potentially affect the modeling of strong lenses \citep{keeton04}, can be taken into account.

Semi-automated algorithms to search for strong galaxy-galaxy lenses are beginning to emerge, for instance in the Canada-France Hawaii Telescope Legacy Survey (CFHTLS) \citep{cabanac07}.  Other arc finding algorithms exist but are currently limited to searching for cluster arcs at space based resolution \citep{lenzen04} \citep{seidel07}.  The development of automated algorithms to search for strong galaxy-galaxy lenses is of interest in order to create large uniform samples in future imaging surveys where selection effects are understood \citep{cabanac07}.  The current generation of weak lensing surveys such as the Deep Lens Survey (DLS) \citep{wittman02} and the CFHTLS can be used as testing grounds for the development of these search techniques.    

Here we present a new method to search for strong galaxy-galaxy lens systems in optical imaging surveys.  Our technique focuses on using the shape of the strongly lensed background galaxy (the object which visually stands out in an image) to search for systems.  This method has the advantage of being independent of lens galaxy type, allowing us to recover systems that are due to massive ellipticals or potentially dark haloes.  Systems with ellipticals as a lens galaxy are the most massive and will produce the highest separation images, making them easier to detect in ground based imaging.  Systems produced by dark haloes should be detectable from the ground, if they do exist.  Since the lens is dark in this case, there are no lens-arc separation issues to deal with.  Excellent dark candidate systems would have a primary arc and secondary arc (with the same color as the primary) at a smaller separation from the lens centre.  In general spiral systems will be harder to detect in ground based imaging since they are lower mass and therefore produce images with lower lens-arc separation.
We apply our technique to the Deep Lens Survey F2 field. 

The outline of the paper is as follows: In $\S \ref{sec:data}$ we discuss the dataset used to apply our search technique.  In $\S \ref{sec:arcpro}$ the properties of arcs are discussed, which are used to define an arc parameter space.  Arcs are detected using a pixel thresholding method described in $\S \ref{sec:detect}$.  In $\S \ref{sec:sims}$ the parameter space of arcs is generated using simulations of strong galaxy-galaxy lensing.  Results of these simulations are discussed in $\S \ref{sec:simresults}$.  In $\S \ref{sec:search}$ we apply our search method to the DLS F2 field and present examples of detected candidates.

Throughout the paper we assume a standard cosmology with $\Omega_{m}=0.3$ and $\Omega_{\Lambda}=0.7$.

\section{Data}
\label{sec:data}

Data used for this study are obtained from the Deep Lens Survey \citep{wittman02} a wide area ($\sim 20$ degrees) multi-band ($BVRz'$) weak lensing survey imaging to $R\sim 26$.  The survey uses the MOSAIC imagers \citep{mosaic} on the Mayall 4-m at Kitt Peak and the Blanco 4-m at Cerro Tololo, Chile.  The survey is spilt into five separate $2^{\circ}\times2^{\circ}$ fields, with each field spilt into a $3\times3$ grid of 40 arcmin $\times$ 40 arcmin subfields.  The $R$ band co-added images total $18,000$ seconds of exposure and the $BVz'$ band total $12,000$ seconds.  Details of the basic data reduction, calibration, and co-addition pipelines are described in \citet{wittman06}.  

In this study we restrict our search to the DLS F2 field (centred on $\alpha=09^{\mathrm{h}}18^{\mathrm{m}}00^{\mathrm{s}}$ $\delta=30^{\circ}00\arcmin00\arcsec$ J2000) which has complete imaging.  For our strong lens search we use only the $R$ band since this is the deepest filter with the best seeing ($<0.9$ arcsec).  The four color filters in the DLS will be used in future weak lensing studies to derive photometric redshifts for source galaxies.  At the time of our study photometric redshifts were not available for the DLS F2 field, and therefore are not used in our analysis.

\section{Search Technique}
\label{sec:arcpro}

The primary motivation for our search technique was to measure what visually stands out to the eye when looking at a strong lensed galaxy, that is the arc shape.  It was first pointed out by Goldberg \& Natarajan (2002) that the shape of an arc is stored in one component of the octopole (or third order) moment.  Their study and subsequent studies \citep{goldberg05} focused on examining the octopole moments in the weak lensing limit.  Here we use this component (along with the other third order moments) to search for arcs in the strong lensing regime. 
  
\subsection{Octopole moments}
\label{sec:oct}

\begin{figure}
\includegraphics[width=84mm]{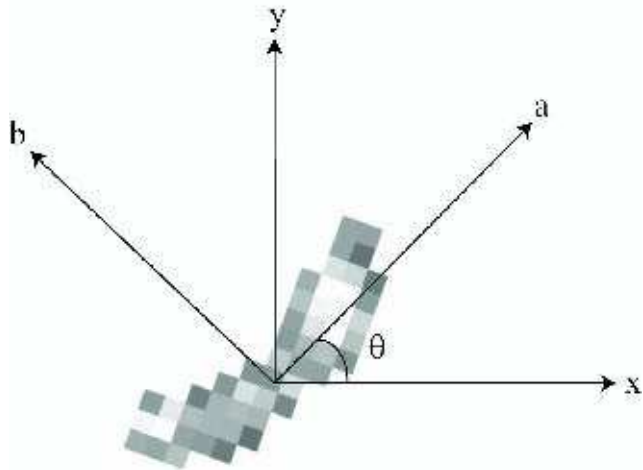}
\caption{The coordinate system in which the arcness of each object is measured.  The image coordinates are given by the (x,y) axes, and (a,b) define the object major and minor axis.  The position angle $(\theta)$ of the object is measured relative to the image x axis.  The arcness of the object is chosen to be measured relative to the major axis of the object.} 
\label{fig:arcgeom}
\end{figure}

Since an arc can have a random orientation in an image, the octopole moments are measured relative to each object instead of the image axis.  For each object we measure the octopole moments relative to the object major axis (Fig. \ref{fig:arcgeom}).  By definition the position angle $(\theta)$ of the object is the angle between the x image axis and the major axis of an object, where
\begin{equation}
\theta=\frac{1}{2}\mathrm{tan}^{-1}(\frac{2I_{xy}}{I_{xx}+I_{yy}}).
\end{equation}
The octopole moments are measured in the frame of each object using the rotation matrix
\begin{equation}
\left(\begin{array}{c}
\Delta x_{i}\\
\Delta y_{i}\\
\end{array}\right)
\left(\begin{array}{cc}
\mathrm{cos}\theta & \mathrm{sin}\theta \\
-\mathrm{sin}\theta & \mathrm{cos}\theta \\
\end{array}\right)
\left(\begin{array}{c}
x_{i}-\overline x\\
y_{i}-\overline y\\
\end{array}\right)
\end{equation}
where $\overline x$ and $\overline y$ define the centroid of the object, and $\theta$ is the position angle.  The four components of the octopole moments are then, $I_{aab}=\frac{\sum_{i} I_{i}(\Delta x_{i})^{2}(\Delta y_{i})}{\sum I_{i}}$, $I_{abb}=\frac{\sum_{i} I_{i}(\Delta x_{i})(\Delta y_{i})^{2}}{\sum I_{i}}$, $I_{aaa}=\frac{\sum_{i} I_{i}(\Delta x_{i})^{3}}{\sum I_{i}}$, $I_{bbb}=\frac{\sum_{i} I_{i}(\Delta y_{i})^{3}}{\sum I_{i}}$, where a and b are the semi-major and minor axis.  The octopole moments are further made unitless and size independent by normalizing by the cube of the object size (size$=\sqrt{I_{xx}+I_{yy}}$).  The quantity $I_{xx}+I_{yy}$ is rotationally invariant but we have measured this in image coordinates.

\begin{figure}
\includegraphics[width=84mm]{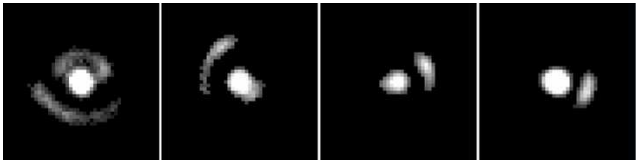}
\caption{\small Examples of simulated strong galaxy-galaxy lens systems with different measured levels of arcness.  In each, the data image is mapped to the segmentation image at the threshold in which the arc is detected.  The foreground lens galaxy is centred on each image, and the arcness is measured on the primary arc in each system.   From left to right the arcness of the primary arc in each system is : 0.20, 0.15, 0.10, 0.05.}
\label{fig:arclevels}
\end{figure}

Relative to the major axis, $I_{aab}$ is the component of the octopole moment that measures the arc shape of the strongly lensed galaxy.  To provide a concavity independent measure of the arc shape we take its absolute value $|I_{aab}|$.  This normalized component is referred to throughout the rest of the paper as the galaxy ``arcness" (given by the symbol : $a$), where  
\begin{equation}
a=\frac{|I_{aab}|}{(I_{xx}+I_{yy})^{1.5}}.
\end{equation}
To better illustrate the arcness of strongly lensed galaxies, examples of simulations with different values of measured arcness are shown in Fig. \ref{fig:arclevels}.  Arcs with a high level of arcness correspond to systems in which the system is significantly tangentially distorted.      

Strongly lensed arcs measured relative to the major axis should also have a small $I_{abb}$ value, making this component of the octopole moment useful as a control. This component is referred to as the ``anti-arcness" (given by the symbol : $\bar{a}$), where

\begin{equation}
\bar{a}=\frac{I_{abb}}{(I_{xx}+I_{yy})^{1.5}}.
\end{equation}

The remaining components of the octopole moments $I_{aaa}$ and $I_{bbb}$, measure the skewness along the major $(s_{a})$ and minor $(s_{b})$ axes respectively
\begin{equation}
s_{a}=\frac{I_{aaa}}{(I_{xx}+I_{yy})^{1.5}}
\end{equation}
 
\begin{equation}
s_{b}=\frac{I_{bbb}}{(I_{xx}+I_{yy})^{1.5}}.
\end{equation}
 
\subsection{Ellipticity}
\label{sec:ellipticity}
In addition to the octopole moments, the ellipticity of a strongly lensed galaxy is also a useful arc property.  The ellipticity is measured in terms of its $e_{1}$ and $e_{2}$ components, where $e_{1}=\frac{I_{xx}-I_{yy}}{I_{xx}+I_{yy}}$, $e_{2}=\frac{2I_{xy}}{I_{xx}+I_{yy}}$, and the total ellipticity (e) is given by $e=\sqrt{e_{1}^{2}+e_{2}^{2}}$.  The quadrapole moments are measured here in the most basic sense, that is $I_{xx}=\frac{\sum_{i}I_{i}(x_{i}-\overline x)^{2}}{\sum_{i} I_{i}}$, $I_{yy}=\frac{\sum_{i} I_{i}(y_{i}-\overline y)^{2}}{\sum_{i} I_{i}}$, and $I_{xy}=\frac{\sum_{i} I_{i}(x_{i}-\overline x)(y_{i}-\overline y)}{\sum_{i} I_{i}}$.  No weights are used in calculating the second order moments in order to avoid making assumptions about the light profile along the arc.

\subsection{Arc S/N}
\label{sec:sn}
Another important property of arcs to consider is their range of expected S/N.  The object S/N is computed from the standard S/N equation using the object flux, the noise per pixel, and the gain ($\simeq 54 e^{-}\rm{pixel^{-1}}$, for the co-added DLS R band image) at each detection threshold.  At some minimum S/N it is expected that the measured arcness becomes unreliable.  The minimum S/N for the DLS is discussed in $\S \ref{sec:minsn}$.

\subsection{Additional parameters}
In addition to the above parameters we also tested others that were not particularly useful.  These included the arc length and arc width, which are typically used to study the shapes of cluster arcs \citep{miralda93}.  At galaxy scales, arcs are not as drastically elongated as arcs produced by clusters, and most of the information contained in these parameters is stored in the ellipticity parameter.

\section{Arc Detection}
\label{sec:detect}
\subsection{Pixel thresholding}
Detecting strongly lensed galaxies (arcs) is challenging since arcs typically occur near a foreground lens galaxy and also can have a complicated surface brightness profile.  Their detection therefore depends on the ability of a detection algorithm to separate the arc from the lens, and also to avoid splitting the arc into multiple objects (and therefore lose its arc shape).  We chose to detect arcs in the DLS using SExtractor \citep{bertin96}, a object detection and shape measurement program.  
SExtractor typically detects objects by splitting a group of connected pixels using a `deblending' algorithm which assigns pixels to each object by computing a probability of belonging to each object.  In the case of separating a lens galaxy and an arc, this step can possibly affect the shape of the resulting arc.  Since there is no way for the user to control this step, we instead chose to use a pixel thresholding approach.  Here the deblending algorithm in SExtractor is turned off and the program is run at a series of detection thresholds.  This is similar to how the SExtractor deblending works, except here the shape information is preserved at each threshold, and only thresholding determines which object a given pixel is assigned.



\subsection{Implementing pixel thresholding}
\label{sec:multi}
 
For a given detection threshold, SExtractor is used to create a segmentation image where the detection threshold is measured relative to the image RMS value.  The segmentation image is then used as input to our custom cataloging software which measures the octopole moments.  Using this global per pixel value of the noise is better than using the SExtractor background map, which is a smoothed map of the image noise.  We additionally made minor modifications to version 2.3.2 of the SExtractor code in order to accept a large number of objects in the segmentation image, which primarily occurs in the DLS fields at low detection thresholds.

\subsection{Thresholds}

The minimum object area is set sufficiently small (below the PSF) in order to avoid rejecting real arcs in the DLS.  The minimum detection threshold to use was determined from 100 realizations of the simulation noise, where the threshold was increased until there was a zero probability of detecting noise as a real object.  This resulted in a minimum detection threshold of $2.5$.  The maximum detection threshold is not as well defined as the minimum, but we set this detection threshold relative to the maximum S/N arc produced by our simulations, which is separated from the lens galaxy at a detection threshold of $20$.

\begin{figure}
\includegraphics[width=84mm]{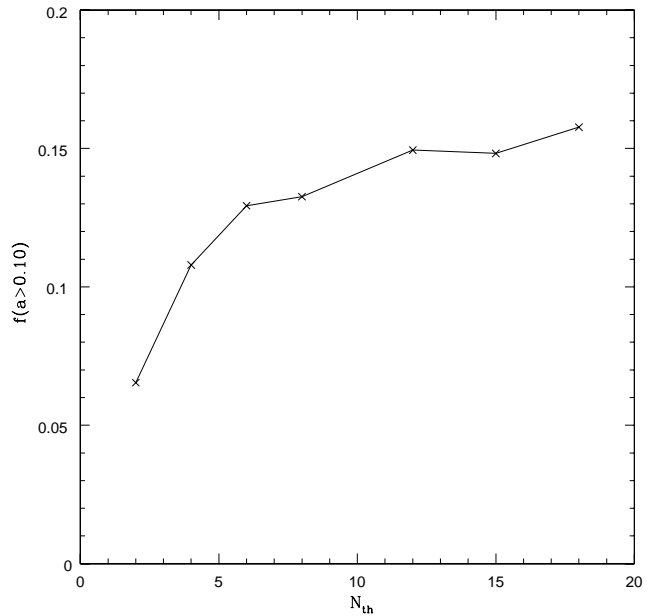}
\caption{The fraction of simulated arcs detected above an arcness level $a>0.10$, as a function of the number of detection thresholds.  As the number of detection thresholds increases, the fraction of recovered simulated arcs flattens at $\sim 15$ thresholds.  This is the number of detection thresholds used to sample the shape of arcs in the DLS.} 
\label{fig:multi}
\end{figure}

Logarithmically spaced thresholds are used from the minimum detection threshold up to the maximum threshold.  As the number of thresholds is increased we expect to reach a value where there is no gain in using more thresholds.  For a large set of simulations we show in Fig. \ref{fig:multi} the fraction of detected simulated arcs with arcness $a>0.10$ as a function of the number of detection thresholds.  The fraction of detections here flattens for $\sim 15$ detection thresholds.  This is the number of thresholds at which the simulations and data are analyzed.  The overall fraction here is low since our simulations produce systems with all values of arcness, which are dominated by systems with a low arcness level.

\section{Strong Lens Simulations}
\label{sec:sims}
To estimate the parameter space of arcs in the DLS we use a large number of simulations $(90,000)$ of strong galaxy-galaxy lensing.  This is a sufficiently large number of simulations to estimate the arc parameter space in the DLS while still being computationally feasible.  To create realistic properties for the arcs in each simulation we use real galaxies from deep HST fields for our sources (as in \citealt{horesh05}), instead of artifical sources. 

\subsection{Source galaxies}
\subsubsection{Source image database}
\label{sec:database}
\begin{figure}
\includegraphics[width=84mm]{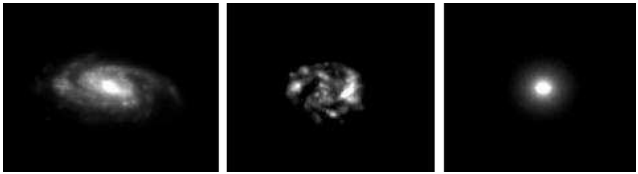}
\caption{Examples of source galaxies in the database used in our simulations of strong galaxy-galaxy lensing (\textit{Left}: Spiral, \textit{Middle}: Irregular, \textit{Right}: Elliptical).  Galaxies were selected from the F606W band of the Hubble Deep Fields (North and South), the Ultra Deep Field, and the Great Observatories Origins Deep Survey.  Each galaxy was thresholded at $4\sigma$ and cleaned to create a ``noise free'' version of the galaxy.}
\label{fig:database}
\end{figure}

To create realistic shapes for strongly lensed galaxies (or arcs) real galaxies from deep HST exposures are used for our sources.  The deepest HST exposures from which to select source galaxies were the Hubble Deep Field North (HDFN) \citep{williams96}, the Hubble Deep Field South (HDFS) \citep{casertano00}, the Great Observatories Origins Deep Survey (GOODS) \citep{giavalisco04}, and the Ultra Deep Field (UDF) \citep{beckwith06}. From each field, galaxies which have the highest S/N were picked from the F606W filter (which matches reasonably well with the DLS $R$ band).  Each galaxy was thresholded at $4\sigma$ (per pixel) and cleaned to create a ``noise free'' image.  This resulted in a database of 90 unique galaxies, 52 from the UDF, 19 from GOODS, 9 from the HDFN, and 10 from the HDFS.  Examples of these galaxies are shown in Fig. \ref{fig:database}.  At the time we began our study morphology was not available for galaxies from the UDF, so galaxies were visually classified into three categories: spirals, ellipticals, and irregulars.  The HDF's did have morphology available, but to be consistent with the UDF sample, Sbc's and Scd's were grouped together as ``spirals''. For each run of the simulation, a source galaxy in the database is selected and rotated to a position angle (both at random), to reflect the random orientation of source galaxies. The source galaxies in the database were also photometrically scaled to the simulated lens galaxy zeropoint. 

\subsubsection{Source image rescaling}

In using real galaxies for sources we are limited by the total number of galaxies with sufficiently high S/N available from the HST fields.  The effective number of available source galaxies is increased by rescaling each initial source galaxy image to reflect the size, magnitude, and redshift distribution as determined from HST surveys.  To rescale the images we initially select a new magnitude for the source galaxy at random from a power law distribution in the range $22<m_{\mathrm{source}}<25$.  The flux in the initial database source galaxy image is then scaled to the new source galaxy magnitude.  A new size is assigned to the source galaxy based on the magnitude-size distribution of galaxies from the UDF.  From the new magnitude and the associated magnitude bin in the UDF catalog, a new size is randomly assigned to the galaxy and the image is scaled to the new size using a linear interpolation routine.  We filtered the initial UDF catalog to include only galaxies where the size is well measured ($R_{1/2}>1.5\times \rm{PSF}$, where the $\rm{PSF}\sim0.06$ arcsec and $R_{1/2}$ is the half-light radius).    A redshift is assigned the source galaxy using the magnitude-redshift distribution of galaxies from the combined photometric redshift catalogs of the HDFN \citep{fernandez99} and HDFS \citep{yahata00}.  The source galaxy is randomly assigned a new redshift based on its initial morphological type and the associated magnitude bin in the redshift catalog.   The initial redshift catalog is separated by galaxy morphological type, and filtered to include galaxies within the range $0.2<z<3.0$.  We chose a maximum source redshift cutoff of $z=3.0$ since few galaxies are expected to be detected above this redshift in the DLS $R$ band at this magnitude and size limit.

\subsection{Lens galaxies}
\label{sec:lens}

Since the lensing cross section is expected to be dominated by ellipticals \citep{fukugita91}, we use elliptical galaxies for our lens galaxies.  For each simulation a lens galaxy magnitude is randomly chosen from a power law distribution in the magnitude range $21<m_{\mathrm{lens}}<23$.  The lens galaxy is assigned a redshift based on the magnitude-redshift relation of elliptical galaxies determined from HST fields (J.A. Tyson 2007, private communication).  Photometric redshifts of lenses in the DLS are not used since these were not available during the course of this study.  Knowledge of the exact lens redshift distribution in the DLS is not crucial since this has only a small effect on the angular diameter distance ratio in equation (\ref{eqn:potential}).  

For each lens galaxy a velocity dispersion is estimated using the Faber-Jackson relation \citep{faber76}.  We consider only galaxies with a velocity dispersion $250 \rm{km} \rm{s^{-1}}<\sigma_{v}<400 \rm{km} \rm{s^{-1}}$, since this is the range which will produce systems with measurable image separations ($>1.0$ arcsec) in the DLS.  To simulate the effects of separating the arc and lens galaxy in real DLS data, an elliptical lens galaxy is included in the final simulated image.

\subsection{Gravitational lens mapping}
\label{sec:mapping}
\begin{figure*}
\begin{minipage}{84mm}
\begin{center}
\includegraphics[width=84mm]{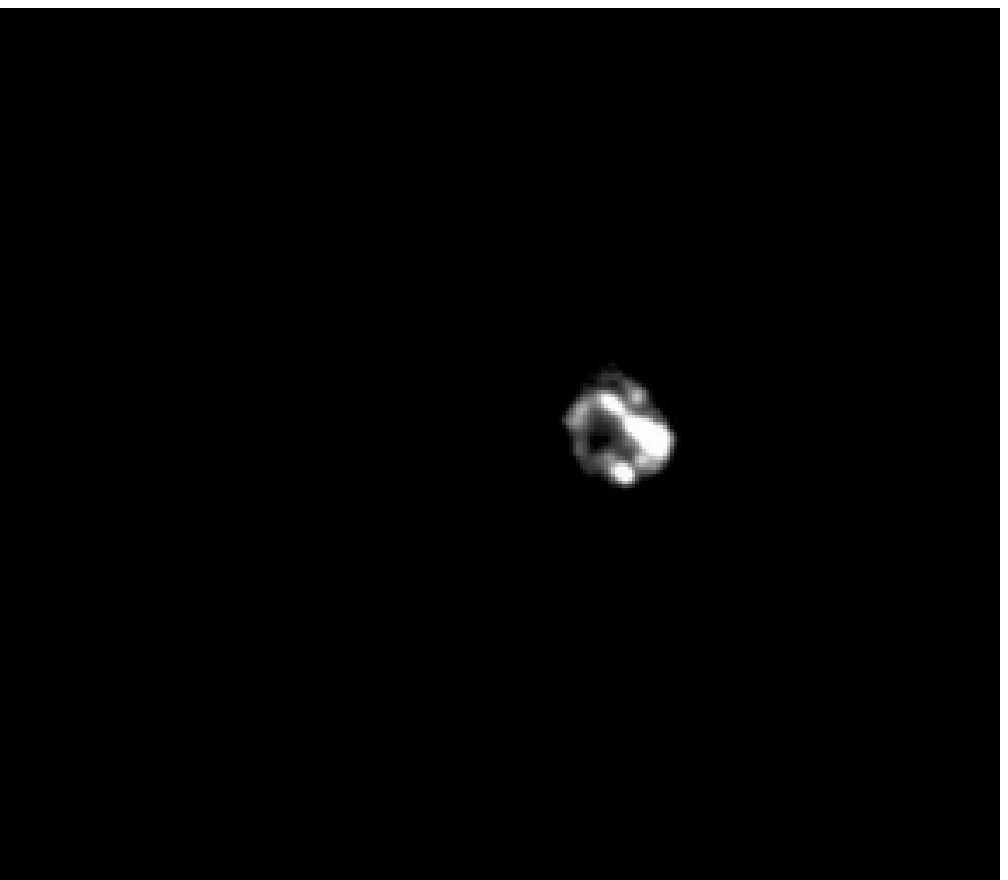}
\end{center}
\end{minipage}
\hfill
\begin{minipage}{84mm}
\begin{center}
\includegraphics[width=84mm]{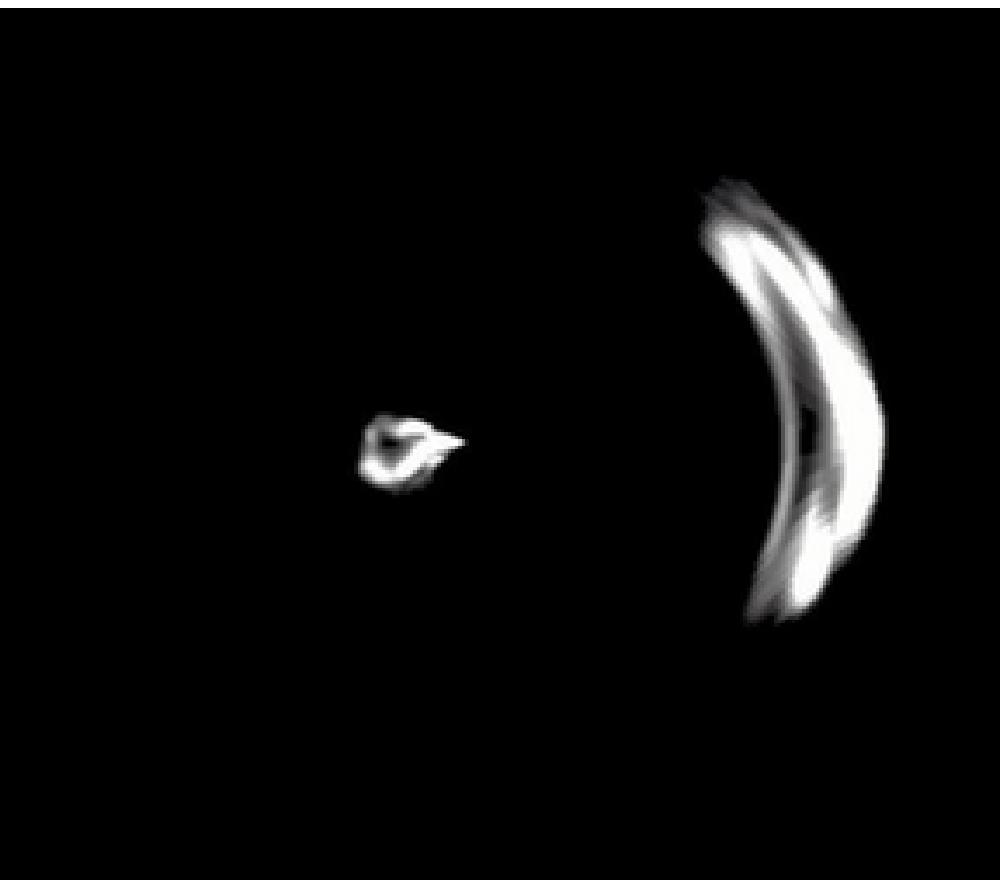}
\end{center}
\end{minipage}
\caption{(\textit{Left}): A galaxy selected from the simulation database is scaled to a new size, magnitude, redshift, and shifted to an impact parameter.  (\textit{Right}): The simulated galaxy on the left is gravitationally lensed by an elliptical potential.  The structure within the source galaxy is evident in both images each of which have a pixel scale of 0.03 arcsec $\mathrm{pixel^{-1}}$.}
\label{fig:lensing}
\end{figure*}

Each source galaxy image is gravitationally lensed using ray tracing via the lens equation $\bmath{\theta}=\bmath{\beta}+\bmath{\alpha}$, where $\bmath{\theta}$ is the image position, $\bmath{\beta}$ is the source position, and $\bmath{\alpha}$ is the deflection given for the projected scalar potential.  The deflection angle $\bmath{\alpha}$ is related to the scalar potential via $\bmath{\alpha}=\bmath{\nabla}\psi$.  To increase the resolution of the lens mapping, each pixel in the lens plane is subdivided into a $4\times4$ grid of sub-pixels.  For the lens potential we chose to use a projected scalar elliptical potential \citep{blandford87}  because of its analytic simplicity and since it generates realistic lensed images.  The deflection for each sub-pixel is calculated from the x and y components of the potential given by \begin{equation}\psi(x,y)=\psi_{o}\sqrt{1+(1-e_{p})x'^{2}+(1+e_{p})y'^{2}}\end{equation} where \begin{equation}\label{eqn:potential}\psi_{o}=4\pi\sigma_{v}^{2}\frac{D_{ls}}{D_{s}}\end{equation} \begin{equation}x'=\frac{x_{\mathrm{lens}}-x}{r_{c}}\end{equation} \begin{equation}y'=\frac{y_{\mathrm{lens}}-y}{r_{c}}.\end{equation}  Here $D_{ls}$ is the angular diameter distance between the lens and source, $D_{s}$ is the angular diameter distance of the source, and $\sigma_{v}$ is the velocity dispersion of the lens. The position of the centre of the lens is given by $x_{\mathrm{lens}}$ and $y_{\mathrm{lens}}$, and $r_{c}$ is the core radius.  Since we are only interested in the primary arc in each system (which is sensitive to the total enclosed mass) we have chosen a small core radius for each system.    The lens galaxy ellipticity ($e_{\mathrm{lens}}$) is coupled to the potential ellipticity ($e_{p}$) via the function \begin{equation}e_{p}=\frac{0.2e_{\mathrm{lens}}}{0.6+e_{\mathrm{lens}}}\end{equation} which constrains the potential ellipticity to the range $0<e_{p}<0.2$, where the surface density remains physical.  Each system is ensured to be strongly lensed by placing the source within the Einstein angle for the system.  An example of lensed source galaxy is shown in Fig. \ref{fig:lensing} (Right).

\subsection{Scaling to the DLS}
\label{sec:scaledls}

\begin{figure*}
\begin{minipage}{84mm}
\begin{center}
\includegraphics[width=84mm]{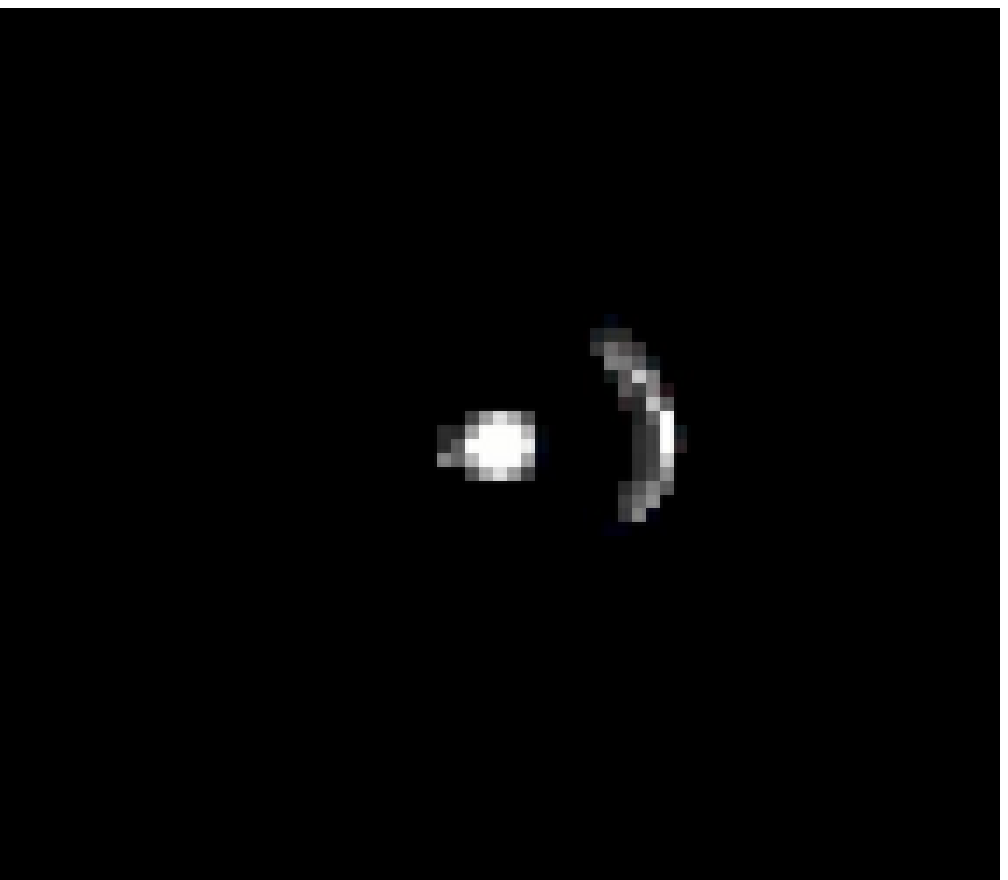}
\end{center}
\end{minipage}
\hfill
\begin{minipage}{84mm}
\begin{center}
\includegraphics[width=84mm]{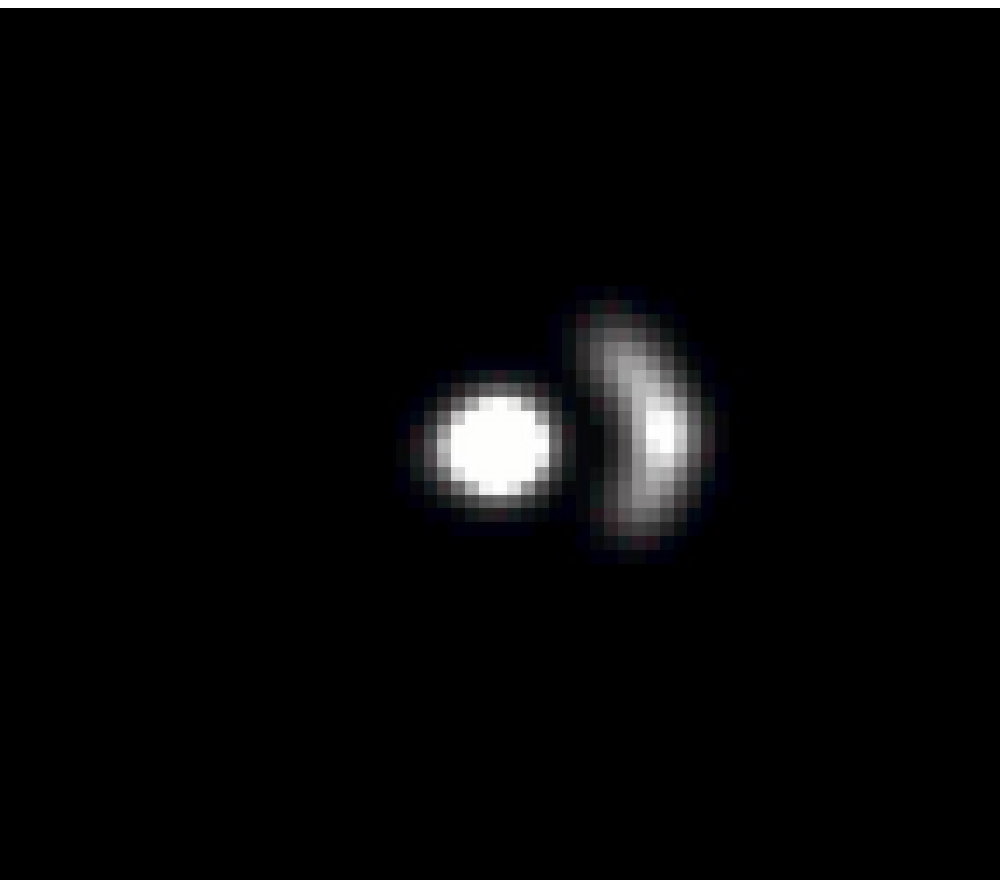}
\end{center}
\end{minipage}
\caption{(\textit{Left}): The image of the gravitationally lensed source galaxy and foreground lens galaxy is scaled to the DLS pixel scale 0.257 arcsec $\mathrm{pixel^{-1}}$ and a foreground lens galaxy is added. (\textit{Right}): The system on the left is convolved with the DLS PSF using a Gaussian with $FWHM=0.9$ arcsec.}
\label{fig:scalings}
\end{figure*}

Each strongly lensed image ($0.03$ arcsec) is scaled to the DLS pixel scale ($0.257$ arcsec) using a linear interpolation routine, where the flux is conserved in the scaling.  The lensed source galaxy image is added to the lens galaxy image to create the strongly lensed system.  An example of a lensed system scaled to the DLS pixel scale is shown in Fig. \ref{fig:scalings} (Left). After scaling to the DLS pixel scale, the image is set to the DLS seeing ($0.9$ arcsec)
by convolving the image with a Gaussian kernel, where the sigma of the Gaussian is related to the FWHM by $\sigma=\frac{\mathrm{FWHM}}{\mathrm{2\sqrt{2ln(2)}}}$.  An example of a lens system smoothed to the DLS seeing is shown in Fig. \ref{fig:scalings} (Right).

To set the S/N of the simulations to the DLS S/N, an object in the simulation (after convolution) was matched to an object in the DLS with the same magnitude and size.  For the matching object in the DLS, the noise $(\sigma_{DLS})$ was measured in the region surrounding the object.  The noise to set in the simulation $(\sigma_{sim})$ is determined by $\sigma_{sim}\simeq \sigma_{\mathrm{DLS}}\frac{F_{\mathrm{sim}}}{F_{\mathrm{DLS}}}$, where $F_{\mathrm{sim}}$ and $F_{\mathrm{DLS}}$ are the flux of the object in the simulation and data respectively.  
 
\begin{figure}
\includegraphics[width=84mm]{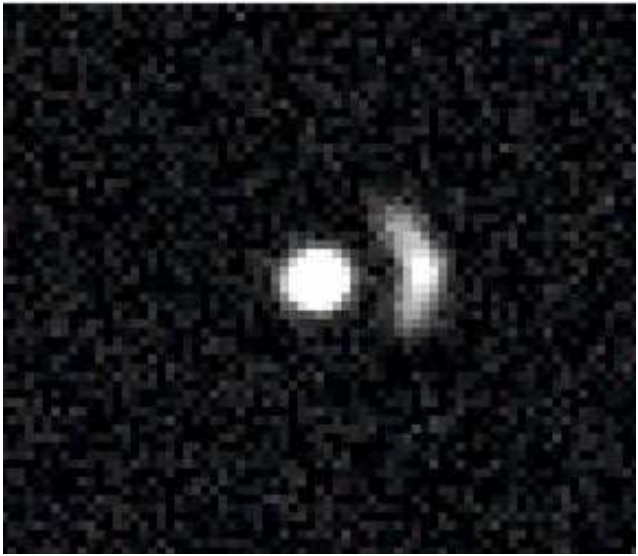}
\caption{A completed simulation of a strong galaxy-galaxy lens system.  In the image noise has been added to the combined image of the foreground lens galaxy and gravitationally lensed source galaxy.  The noise scales the system to the correct DLS S/N.  Most of the structure within the DLS scaled arc is gone, except for the central region of the source galaxy.}
\label{fig:simexample}
\end{figure}

For each simulation Poisson noise is generated on a background (B) which is set by \begin{equation}B=\frac{(\sigma_{\mathrm{sim}} \times \rm{gain})^{2}}{\rm{gain}}\end{equation} where a new random seed is used for each simulation.  For the gain we use the same value as in \S \ref{sec:sn}.  The background value is subtracted from the noise image to create the final noise image.  A simulated strong galaxy-galaxy lens system with noise is shown in Fig. \ref{fig:simexample}.

\section{Simulation Results}
\label{sec:simresults}
\subsection{Detecting simulated arcs}
\label{sec:detectsims}
We use an automated method of extracting arcs from the simulations since a large number of simulations are generated ($\sim90000$).  Arcs are detected in the simulations at exactly the same thresholds that are used for detection on the DLS data.  Since the thresholded catalogs for each simulation contain a galaxy lens and at least one arc, a minimum of two objects is required to be detected in each catalog.  Objects are searched for which have a minimum separation of $1.0$ arcsec from the centre of the lens and are resolved ($\sqrt{2}\times\mathrm{size}>1.00$ arcsec).  Detected objects meeting these criteria in each thresholded simulation catalog are considered arcs.  For a given simulation, if an arc is detected at multiple thresholds one of the thresholds is randomly picked as the shape of the arc and used for the result of the simulation.  

\subsection{Arc parameter distributions}
\label{sec:simoct}
\begin{figure}
\includegraphics[width=84mm]{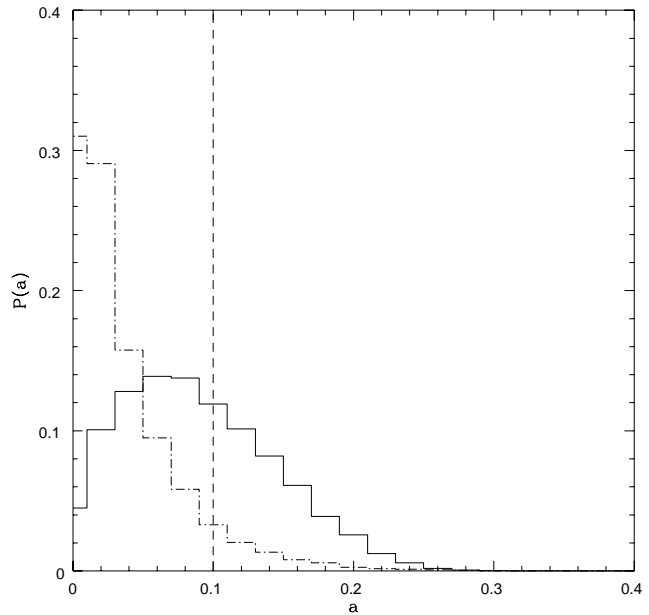} 
\caption{The normalized distribution of arcness for simulations (\textit{solid}) and for data in a DLS subfield (\textit{dot-dash}).  Both histograms show objects which are extended ($e>0.6$) and are above the minimum $S/N$ level ($S/N>40$).  Most extended objects in a DLS subfield have a small arcness with the highest fraction of arcs at $a\sim 0.02$.  The most probable arcness for simulated arcs occurs at a higher arcness level $a\sim 0.06$.  The dashed line at $a=0.1$ represents a possible cut to make in the simulations above which to select arcs which are significantly distorted.}
\label{fig:arcdist}
\end{figure}

The detection method described in $\S \ref{sec:detectsims}$ was used to determine the properties of our simulated arcs.  In Fig. \ref{fig:arcdist} the normalized distribution of simulated arcness is shown (solid line), where the peak in the distribution occurs for a small value of arcness.  The low end of arcness represents source galaxies that have been strongly lensed, but where the galaxy shape is not particularly curved.  Data for a single subfield is also shown (dot-dash line) where the peak in arcness for data occurs at a lower value of arcness.  The offset between the simulations and data indicates that there is discrimination power in the arcness parameter, between arcs and other objects in the subfield.  The dashed line indicated at arcness value of 0.10 represents a possible cut in arcness above which to select arcs which are significantly distorted. 

\begin{figure}
\includegraphics[width=84mm]{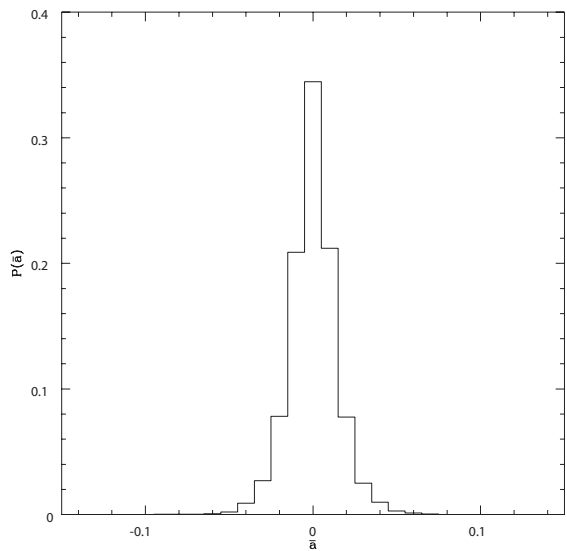}
\caption{This normalized distribution of anti-arcness ($\bar{a}$) for simulations which have been significantly distorted $(a>0.10)$.  The anti-arcness values are peaked around zero, which indicates that arcs which are significantly distorted relative to their major axes have a small distortion relative to their minor axis.}
\label{fig:antiarcness}
\end{figure}

In Fig. \ref{fig:antiarcness} the distribution of anti-arcness is shown for arcs with an $a>0.10$.  These highly distorted arcs are distributed tightly around zero in this parameter space.  This is because arcs which are significantly distorted relative to the major axis are not significantly distorted relative to their minor axis.  The resulting distributions for the remaining octopole moments, the skewness along the major and minor axes, are similarly distributed about a zero value.  Both distributions are Gaussian with a FWHM$\simeq0.8$ for $s_{a}$ and FWHM$\simeq0.03$ for $s_{b}$, and therefore the skewness along the major axis varies significantly more than the skewness along the minor axis.

\begin{figure}
\includegraphics[width=84mm]{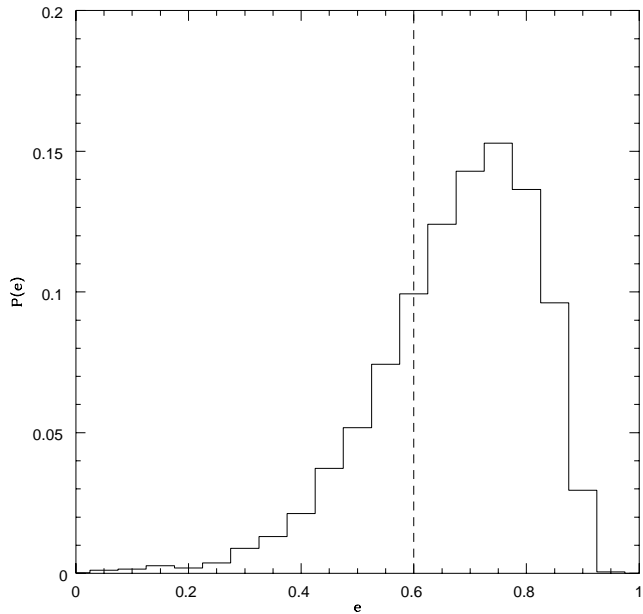}
\caption{The distribution of ellipticity for arcs which are significantly distorted ($a>0.10$) and which meet the minimum $S/N$ ($S/N>40$).  The most probable ellipticity occurs for arcs with ($e\sim 0.75$).  The dashed line at an ellipticity of $e=0.6$ indicates a cut to make in ellipticity, above which to select arcs that are extended.}
\label{fig:ellipticity}
\end{figure}

In Fig. \ref{fig:ellipticity} the distribution of ellipticity is shown for simulated arcs with $a>0.10$.  The majority of significantly distorted arcs occur at high ellipticity $e>0.6$, where a line at this ellipticity is shown to represent a cut above which to select arcs that are extended.    

\subsection{S/N}
\label{sec:minsn}
\begin{figure*}
\begin{minipage}{84mm}
\begin{center}
\includegraphics[width=84mm]{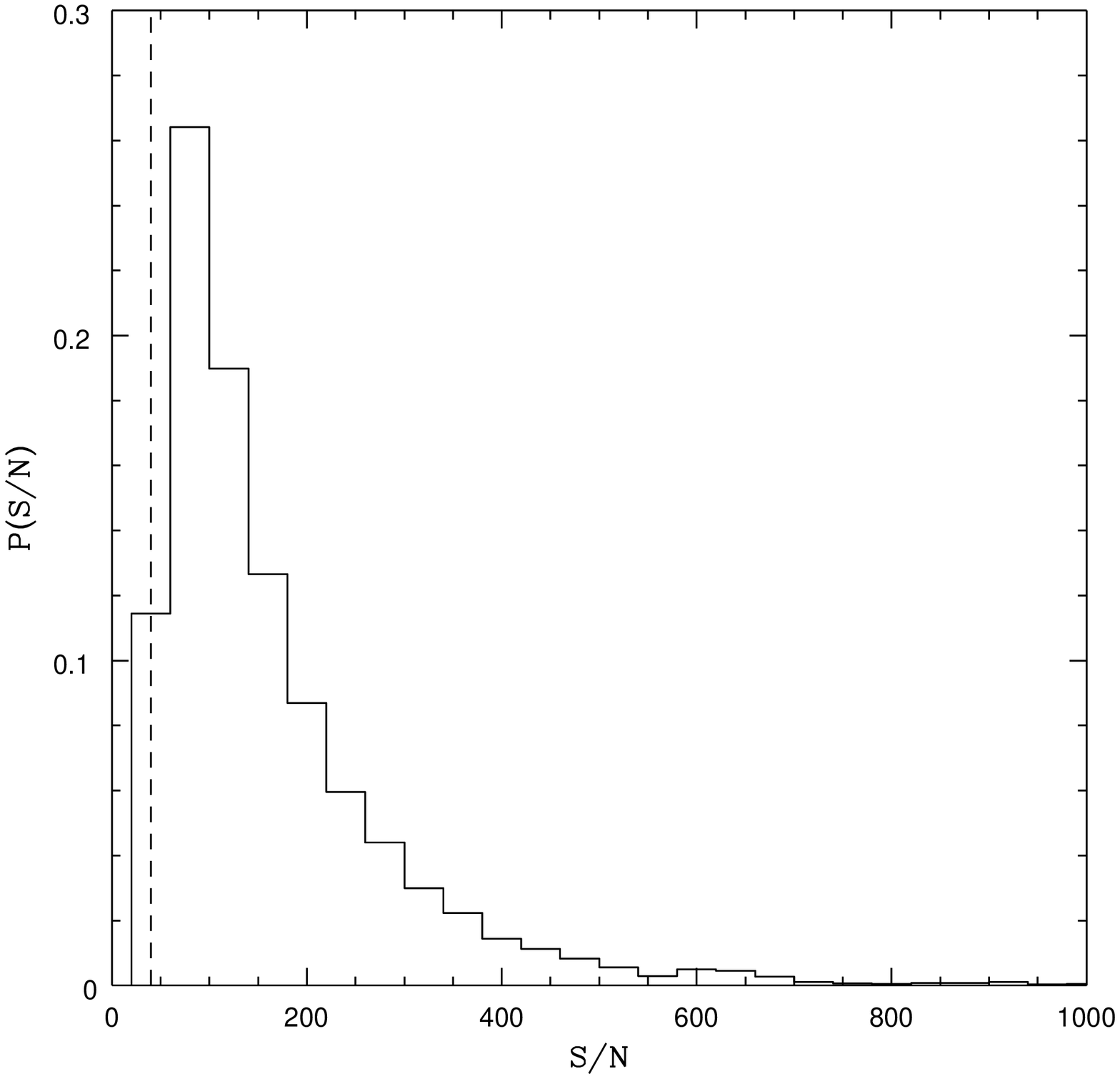}
\end{center}
\end{minipage}
\hfill
\begin{minipage}{84mm}
\begin{center}
\includegraphics[width=84mm]{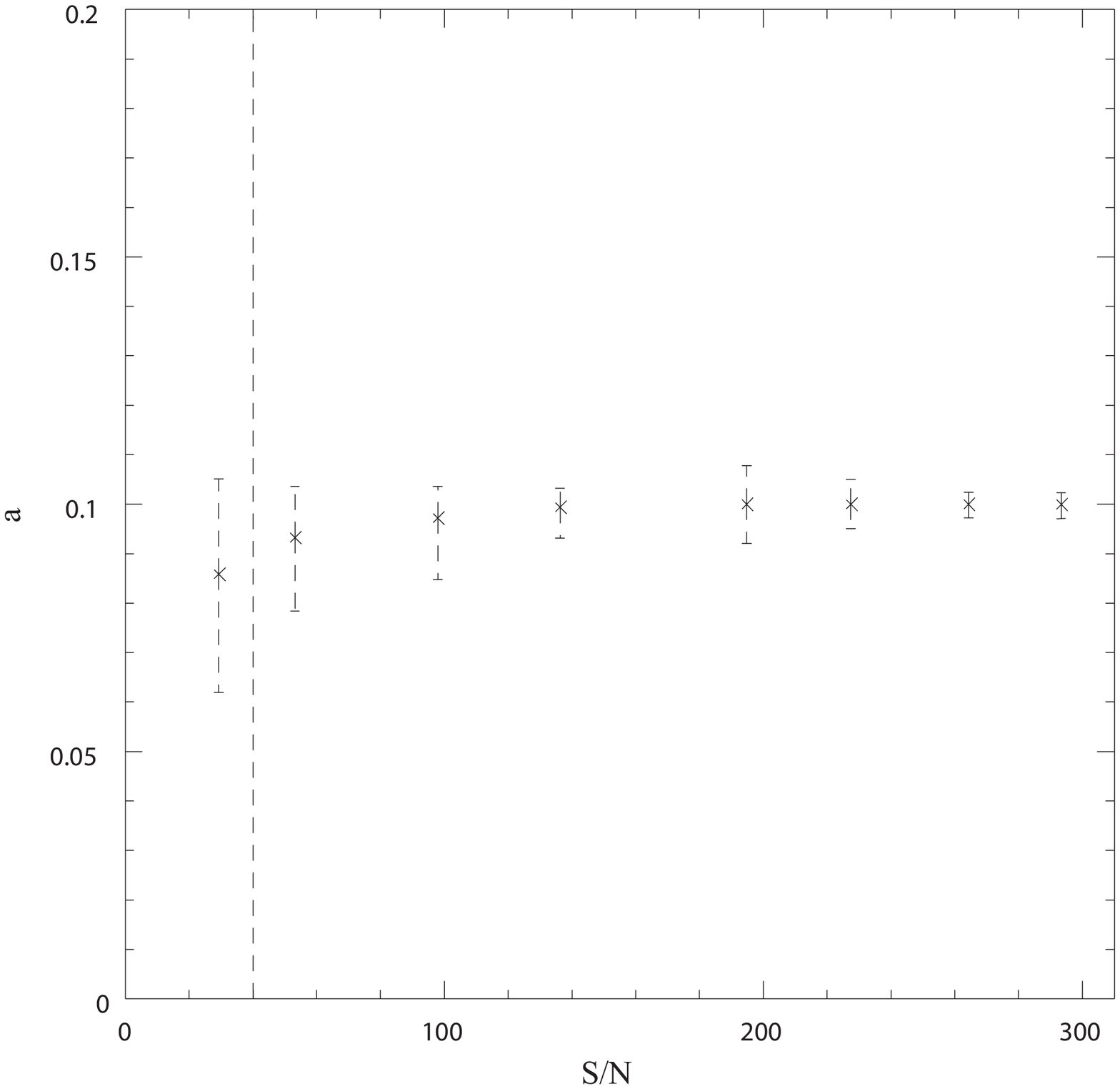}
\end{center}
\end{minipage}
\caption{(\textit{Left}): The S/N range of simulated galaxy-scale arcs in the DLS.  The dotted line at a S/N of 40 indicates a minimum value of S/N that an arc should have in order for the arcness to be considered reliable.  The most probable S/N occurs at $S/N\sim 70$ with virtually all simulated arcs having a $S/N<1000$.(\textit{Right}): A set of simulated arcs is set to different background levels and scaled to an arcness of 0.10.  Each point is the median value within each $S/N$ bin and error bars are set to the $20\%$ and $80\%$ values within each bin.  The uncertainty in arcness becomes large at a $S/N\sim40$ shown as a dotted line.}
\label{fig:sntest}
\end{figure*}

In Fig. \ref{fig:sntest} (Left) the distribution in S/N for arcs with an $a>0.10$ is shown.  The most probable S/N occurs at $S/N\sim70$ with virtually all simulated arcs having a $S/N<1000$.  The minimum simulated arc S/N extends to very small values of S/N, however we expect that at some S/N level the
measured arcness becomes unreliable.  To determine this minimum S/N, a set of simulated arcs were smoothed to seeing of $0.9$ arcsec, and the arcness for each arc was measured at a series of different background levels.  For this bootstrap method, 100 realizations of the noise were used at each background level.  The results of this test are shown in Fig. \ref{fig:sntest} (Right).  To compare arcs from different simulations, the arcness at all S/N levels are scaled relative to the highest S/N level (which has been scaled to an arcness of 0.10).  Measurements at different noise realizations for the arc are then grouped into different bins of S/N.  Each point in the plot represents the median value for all simulations within that S/N bin.  The error bars are relative to the arcness values at $20\%$ (lower error bar) and $80\%$ (upper error bar) within each bin.  At high S/N levels the arcness for the arcs remains relatively stable, but the uncertainty in arcness becomes high for simulations with $S/N \le 40$.  Therefore we set our minimum S/N to reliably measure the arcness at $S/N=40$, indicated by a line in Fig. \ref{fig:sntest} (Left).  We could have been more restrictive and raised the minimum $S/N$ here, but did not want to reject a large number of potential strong lens systems with this cut.

\subsection{Lensing fraction}
\label{sec:frac}

Our simulations were generated by placing the source galaxy at random positions within the Einstein radius of the system.  Probing the entire space of possible arc systems allows us to determine the fraction of systems that will produce a certain arc configuration.  This fraction is shown in Table \ref{tab:lenstab} for the different ranges of arcness used in $\S \ref{sec:search}$. The fraction represents the number of simulations which are above the arcness level, extended $(e>0.6)$, and meet the minimum $S/N$ level $(S/N>40)$.  It is normalized by the total number of simulations which are considered strongly lensed and meet our minimum lens-arc separation ($>1.00$ arcsec).  

\begin{table}
\caption{Fraction of Simulations With Different Arcness Levels} 
\begin{tabular}{@{}lcc}
\hline
arcness range & f $(\%)$\\
\hline
$0.05-0.10$ & 15.4\\
$0.10-0.15$ & 10.9\\
$>0.15$ &  6.7\\
\hline
\end{tabular}
\label{tab:lenstab}
\end{table} 

Arcs which are the most significantly distorted correspond to only a small fraction $\sim 7\%$ of the possible strongly lensed systems in the DLS; however as the arcness level is reduced more of the parameter space of arcs is recovered.  For example, including all arcs with $a>0.10$ would correspond to $\sim 18\%$ of the possible parameter space, or including all arcs with $a>0.05$ would correspond to $33\%$ of the parameter space.

\section{Arc Search}
\label{sec:search}
\subsection{Appication to DLS data}
We searched for arcs in the DLS F2 field using the arc parameters outlined in $\S \ref{sec:arcpro}$ independent of the position of any foreground lens galaxy.  Each DLS F2 subfield is cataloged using the pixel thresholding method described in $\S \ref{sec:detect}$.  In each thresholded catalog definite spurious objects occur near subfield edges and bright stellar haloes.   These regions are filtered by creating SAOImage DS9 region files by hand.  Using the probability selection described in $\S \ref{sec:prob}$ we performed a blind search for systems in the DLS F2 field $\S \ref{sec:blind}$.  Searching the entire parameter space of arcs produced a large number of spurious detections $\sim 40,000$ per square degree.  This is because all objects which are strongly lensed are searched for including those with a small shape distortion.  To reduce the number of spurious detections we restricted the type of arc to search for, searching only for galaxies which are significantly distorted and extended.  This corresponds to searching for objects which have a high arcness and high ellipticity.  The space of highly distorted and extended arcs is generated by selecting only these systems from simulations.    

\begin{figure*}
\begin{minipage}{84mm}
\begin{center}
\includegraphics[width=84mm]{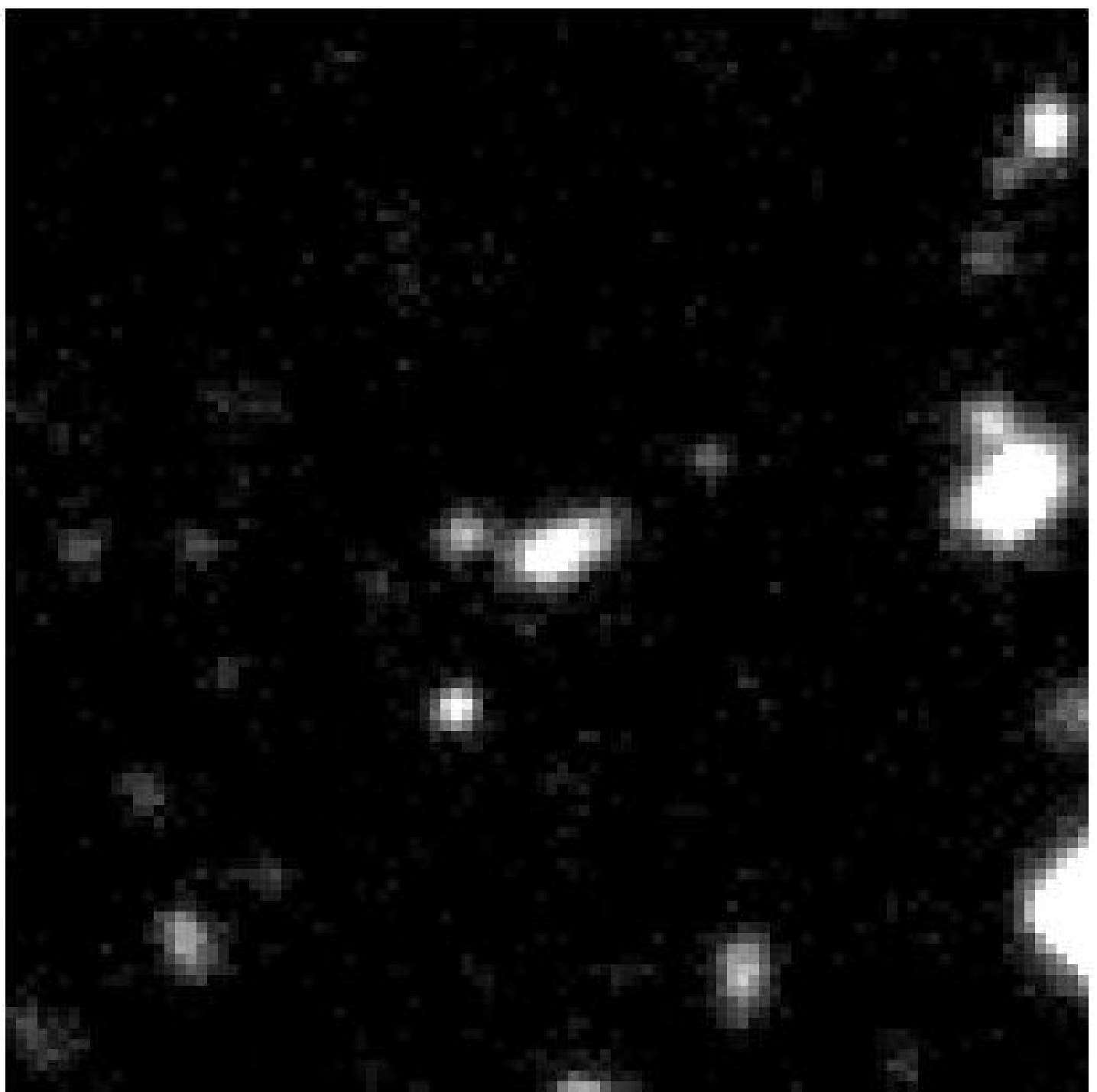}
\end{center}
\end{minipage}
\hfill
\begin{minipage}{84mm}
\begin{center}
\includegraphics[width=84mm]{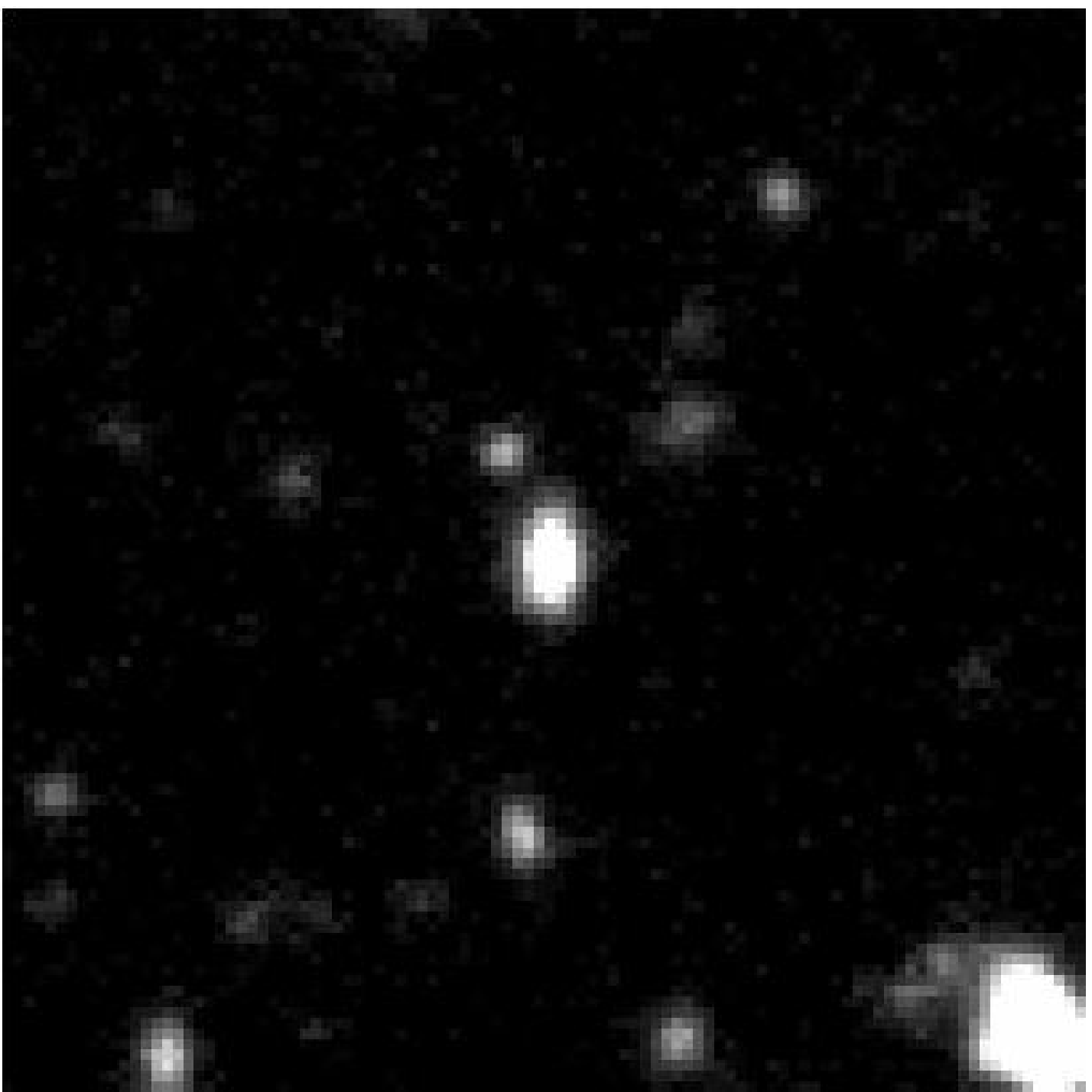}
\end{center}
\end{minipage}
\caption{Examples of typical spurious detections rejected during visual inspection.  Centred on each image are objects which lie in the parameter space with $a>0.15$ and $e>0.6$.  These types of objects are rejected upon visual inspection since they are clear superpositions of two galaxies.  The dimensions of the above images are 26 arcsec $\times$ 26 arcsec.}
\label{fig:spurious}
\end{figure*}

Our blind search for high arcness systems produced a list of candidates systems in the DLS F2 which were visually inspected.  Here objects which were obvious superpositions of two objects were rejected.  We also rejected objects which were clearly spiral arms of a galaxy or were clearly associated with another foreground object.  Examples of typical spurious detections are shown in Fig. $\ref{fig:spurious}$.  


\begin{figure*}
\begin{minipage}{84mm}
\begin{center}
\includegraphics[width=84mm]{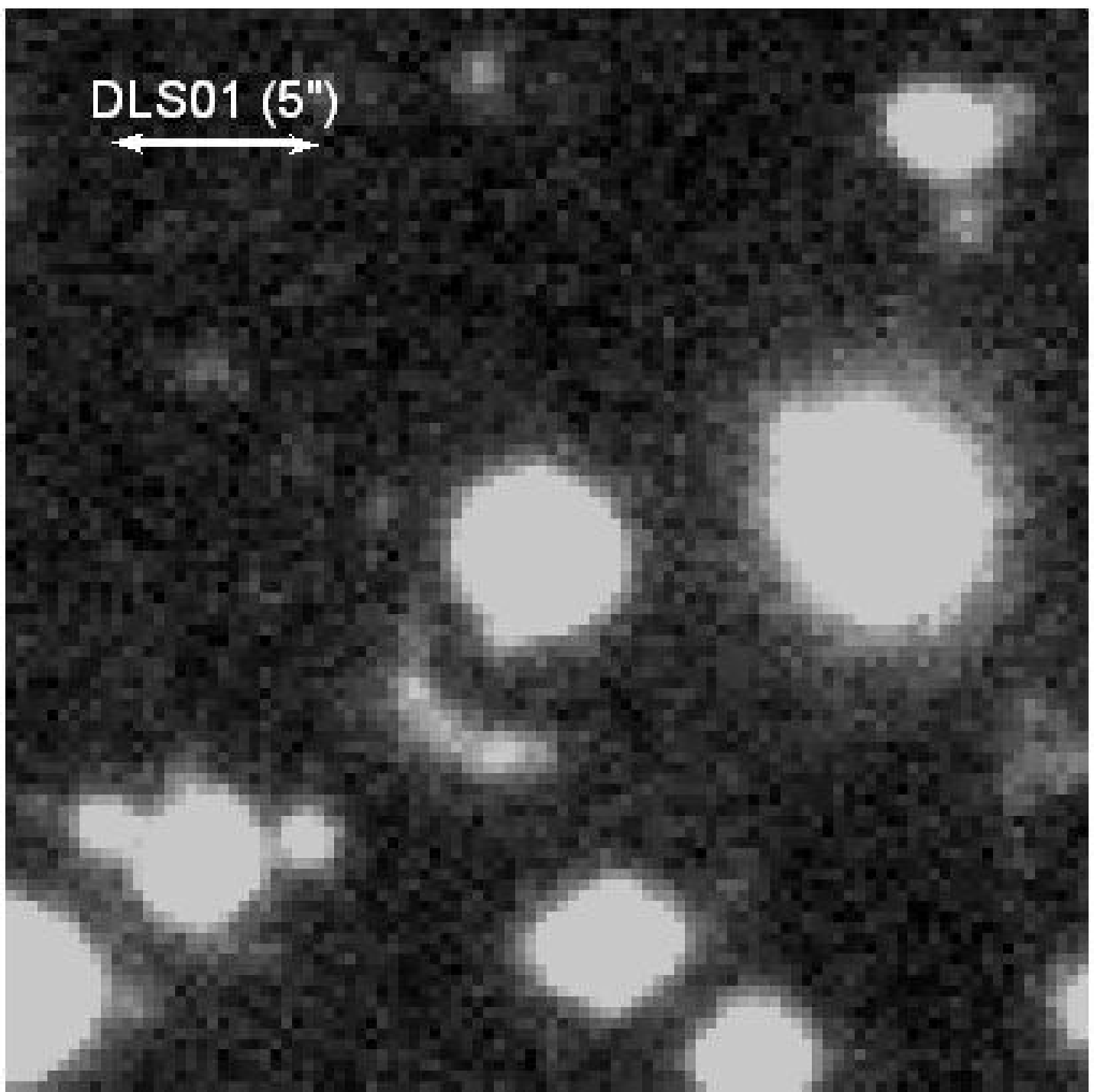}
\end{center}
\end{minipage}
\hfill
\begin{minipage}{84mm}
\begin{center}
\includegraphics[width=84mm]{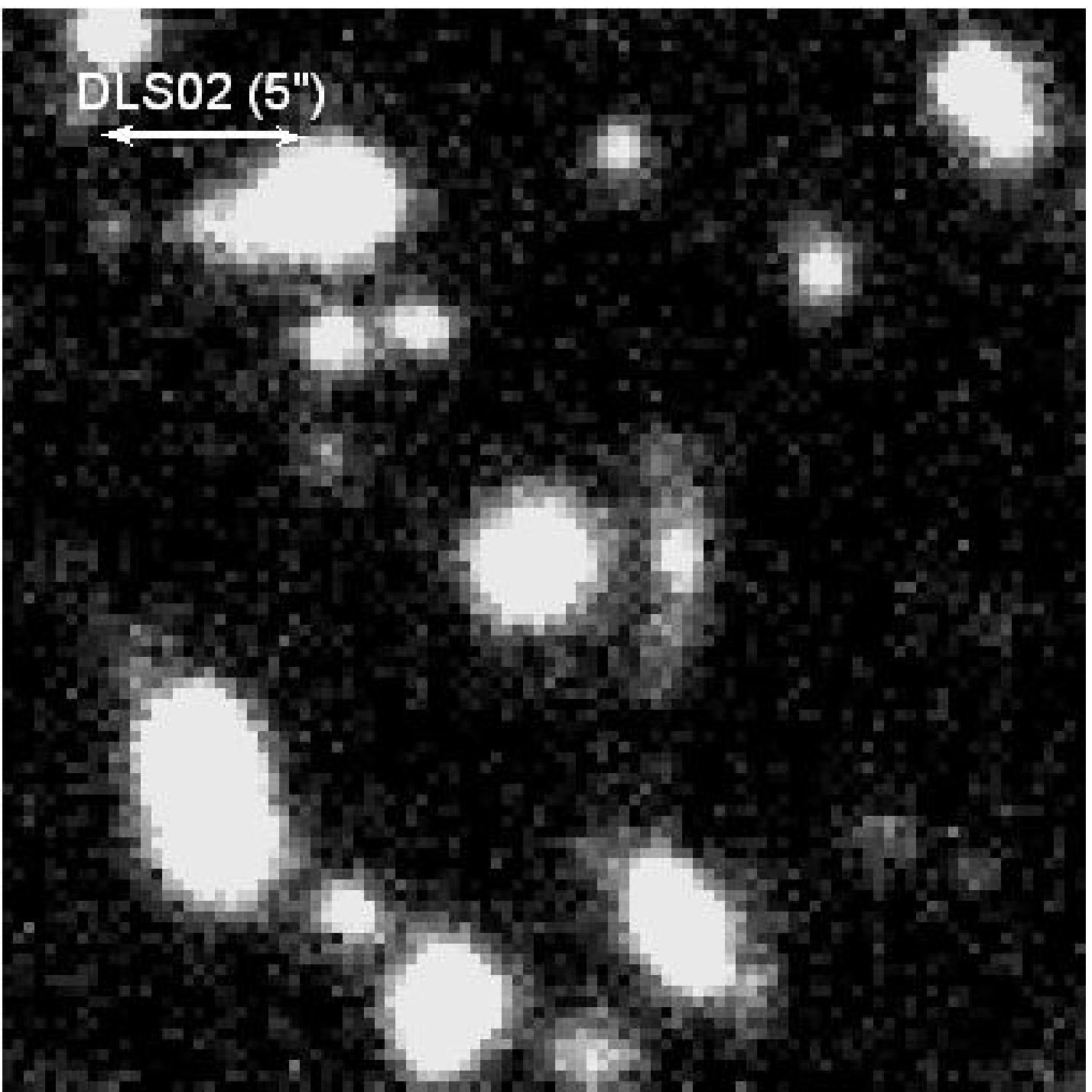}
\end{center}
\end{minipage}
\caption{Examples of arc candidates detected in the DLS F2 field.  The scale of each image is 26 arcsec $\times$ 26 arcsec and the orientation is East is down, North is to the Left.  The candidate on the (\textit{Left}) DLS01 was detected in the parameter space $a>0.15$ and $e>0.6$.  The candidate on the (\textit{Right}) DLS02 was discovered in the parameter space $0.10<a<0.15$ and $e>0.6$. Both arc candidates appear centred towards a corresponding elliptical lens galaxy.}
\label{fig:arccands}
\end{figure*}

\subsection{Probability selection}
\label{sec:prob}
Using the simulation results for each arc parameter in $\S \ref{sec:simoct}$ and $\S \ref{sec:minsn}$ we calculated the probability that an object in the DLS is consistent with the properties of a strongly lensed arc.  We desired a method which was general enough so that any new parameters could be added relatively easily.  This ruled out a Bayesian approach because many of our arc parameters are dependent on each other, and adding a new parameter would become cumbersome.  The method chosen to calculate the probability was based on the occupation probability in the multi-parameter space defined by our simulations.  

The number of arc parameters and the number of bins for each parameter are used to define the multi-parameter space.  The catalog of simulated arc shapes is used to initially populate the space and each bin is normalized in order to assign a bin probability. The simulation probability bins are sorted in descending order while keeping track of the location of each bin.  We additionally define a ``probability contour'' for the space by summing from the maximum probability until a cut in total probability is met.  This probability contour can be used to reject large portions of the parameter space of arcs, however in general we keep the probability high in order to detect a large number of arcs.  The bin sizes are made small in order to maximize the rejection of spurious objects.  To ensure that the bin sizes are not too small (and thus have a fragmented space) the grid in the multi-parameter space was required to be connected.  A given parameter bin must be connected to the other simulation bins in at least one direction along any parameter. 

\subsection{Blind search}
\label{sec:blind}

In our blind search, arcs with high ellipticity were searched for at two different levels of arcness.  The first level was for arcs with $a>0.15$ and $e>0.6$.  Restricting the parameter space here reduced considerably the number of spurious detections, to an average of $\sim 410$ per square degree eliminating $99.9\%$ of all objects in a DLS subfield.  One interesting candidate (DLS01) is discovered at this arcness level shown on the Left in Fig. \ref{fig:arccands} (North is to the left, East is down).  The candidate arc has a measured arcness $a=0.18$, ellipticity $e=0.73$, and magnitude $R=23.5$.  The arc also appears centred relative to an elliptical lens galaxy located southwest of the arc with magnitude $R=20.37$.

We next searched for arcs in the parameter space with $0.10<a<0.15$ and $e>0.6$.  Within this range of arcness the number of spurious detections increases to $\sim1100$ per square degree.  Another interesting arc candidate (DLS02) is detected in this lower arcness space shown on the Right in Fig. \ref{fig:arccands} (North is to the left, East is down).  
This arc candidate has a measured arcness $a=0.13$, ellipticity $e=0.80$, and magnitude $R=23.9$.  The arc also appears centred on an elliptical lens galaxy located directly north of the arc with magnitude $R=21.96$.  

Using our previously discussed criteria for an excellent candidate dark lens in $\S \ref{sec:introduction}$, we find no candidate dark lens systems in the DLS F2 for either parameter space.


\subsection{Visual search}
\label{sec:discussion}
\begin{figure}
\includegraphics[width=84mm]{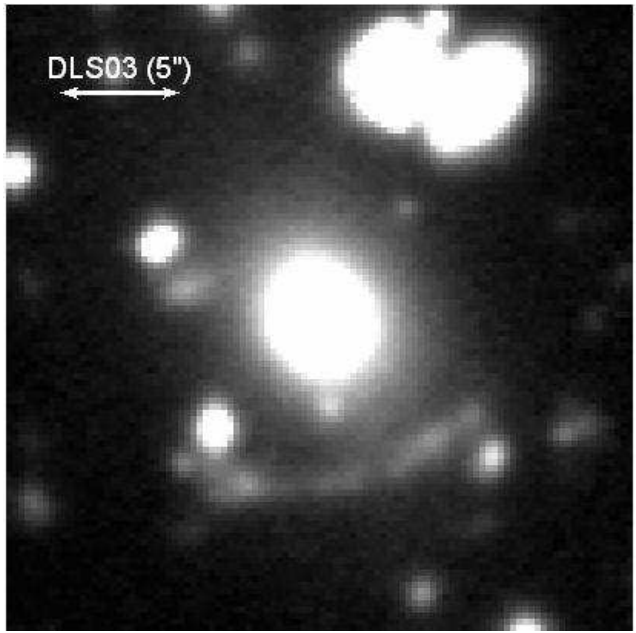}
\caption{The arc candidate DLS03 which is visually detected in the DLS F2 field.  The scale of each image is 26 arcsec $\times$ 26 arcsec and the orientation is East is down, North is to the Left.  This system is not detected using our search technique because of the bright halo of the foreground lens galaxy ($R=18.19$) in the DLS imaging.}
\label{fig:byeeye}
\end{figure}

In addition to our blind search we also performed a visual search in the DLS F2.  Here we viewed each subfield image at high contrast in order to look for arcs within bright lens galaxy haloes that our previous search could have missed.  We recovered one arc candidate (DLS03) in F2 shown in Fig. \ref{fig:byeeye}.  The system was missed in our blind search because pixel thresholding was not able to successfully separate the arc from the foreground lens galaxy at any threshold.  This was due to the brightness of the foreground lens galaxy ($R=18.19$) and the extent of its halo in the DLS imaging.

\subsection{Candidate discussion}

Properties for each candidate strong lens system are given in Table \ref{tab:cands}.  The candidate system DLS01 has a large estimated Einstein radius ($4.9$ arcsec) which is likely due to its proximity to nearby galaxy clusters.  The preliminary weak lensing convergence map for the DLS F2 field (Wittman et. al 2006) reveals a large structure with three overlapping peaks.  A redshift survey of this field has identified these peaks to be Abell 781, CXOU J092053+3029880, and CXOU J092110+302751 \citep{geller05}.  Relative to the centre of the peak in the convergence map DLS01 lies $2.9$ arcmin away.  The visually detected candidate DLS03 also lies close to these clusters ($7.4$ arcmin from the centre) and has the largest estimated Einstein radius ($6.7$ arcsec) in our sample.  The remaining candidate DLS02 has the smallest Einstein radius ($3.4$ arcsec) and is separated by at least $20$ arcmin from any peak in the convergence map. 

We have started follow-up of the visually detected system but leave a detailed discussion and lensing analysis for a future paper.  Redshifts of the lens galaxies in the other systems are unknown, though future photometric redshifts in the DLS F2 will provide estimates.  Spectroscopic follow-up will allow for definitive confirmation of these systems, however since the lens and arc in both of these systems are faint this will require follow-up on 8-m class telescopes. 

It is possible that the candidate arcs in DLS01 and DLS02 could be galaxies which are not strongly lensed but instead have a high intrinsic arcness, for instance due to a tidal interaction.  Follow-up spectroscopy of our candidates would be needed to further examine this possibility.  The frequency of objects with high intrinsic arcness in the DLS or any optical imaging survey is not understood, and warrants future study.  

Additional candidate systems with smaller lens-arc separation could potentially be recovered by restricting our search to regions near foreground lens galaxies and subtracting out the lens galaxy light profile.  For computational reasons we did not pursue this in our current study but this will be explored in future work.


\begin{table*}
\begin{minipage}{128mm}
\caption{Candidate Strong Lens Galaxies in the DLS F2 Field.  The Einstein radius is estimated from the R band data assuming the arc position traces the critical curve.  Arcness and ellipticity parameters are not reported for the candidate DLS03 since this candidate was visually detected.} 
\begin{tabular}{@{}lcccccccc}
\hline
Name & R.A.(J2000) & DEC.(J2000) & $\mathrm{R_{lens}}$ & $\mathrm{R_{arc}}$ & $\mathrm{\theta_{E_{est.}}(\mathrm{arcsec})}$ & arcness & ellipticity\\
\hline
DLS01 & 09:20:19.78 & +30:28:31.81 & 20.37 & 23.5 & $4.9$ & 0.18 & 0.73\\ 
DLS02 & 09:16:29.71 & +29:53:40.46 & 21.96 & 23.9 & $3.4$ & 0.13 & 0.80\\
DLS03 & 09:19:35.06 & +30:31:56.63 & 18.19 & 21.1 & $6.7$ & $N/A$ & $N/A$\\ 
\hline
\end{tabular}
\label{tab:cands}
\end{minipage}
\end{table*}

\section{Conclusion}
A new method of searching for strong galaxy-galaxy lens systems in optical imaging surveys is presented.  Our search method uses the third order moments of galaxies (along with the other arc parameters we have outlined) in order to identify candidate strong lens systems.  Since detection depends only on the properties of the lensed source galaxy, this method can potentially detect systems independent of lens galaxy type, including systems produced by dark lenses.  This method was applied to the Deep Lens Survey, a current wide field ground based optical survey.  Within this dataset our method is sensitive to more massive systems where the lens-arc separation is $>1.0$ arcsec.  This method is particularly successful on systems in which the foreground lens galaxy is faint, which potentially lie at higher redshift.  Systems near bright foreground lens galaxies (which likely lie at lower redshift) are harder to detect with this method, but are recoverable within a survey of this size by viewing the co-added images at high contrast.  In the future other arc detection methods, instead of the pixel thresholding method used here, could help to improve the automated detection of arcs near bright lens galaxies.

Our method should be useful for upcoming wide area ground based surveys such as the DES or future space based imaging surveys such as SNAP (Supernova Acceleration Probe) \citep{aldering05}.  Imaging in the DES will be shallower than the DLS, decreasing the number of spurious detection per square degree.  Here the combination of wide area and good seeing should provide a dataset where a large number of strong galaxy-galaxy lens systems system could be detected.  A significant number of systems could potentially be detected with SNAP, in particular small separation systems ($<1\arcsec$), given the small space based point spread function.  Current HST archival data would provide a good testing ground for our technique at spaced based resolution and should be explored.

\section*{Acknowledgments}
We thank Hossein Khiabanian for many useful comments throughout the course of this work.  We also thank the DLS Collaboration for the reduction and calibration of the DLS data, as well as NOAO for allocating time to the survey.  This work was supported by NSF grant 0134753.  J.M. Kubo was also supported in part by the NASA Rhode Island Space Grant.  IRAF is distributed by the National Optical Astronomy Observatories, which are operated by the Association of Universities for Research in Astronomy, Inc., under cooperative agreement with the National Science Foundation.  This research has made use of SAOImage DS9, developed by Smithsonian Astrophysical Observatory.  Kitt Peak National Observatory, National Optical Astronomy Observatory, is operated by the Association of Universities for Research in Astronomy, Inc. (AURA) under cooperative agreement with the National Science Foundation.


\label{lastpage}

\end{document}